\documentclass[%
aip, 
apl,
preprint,
superscriptaddress,
 amsmath,amssymb,
floatfix,
]{revtex4-1}

\usepackage{graphicx}
\usepackage{dcolumn}
\usepackage{bm}
\usepackage{color}

\begin{document}


\title{Near-field energy transfer between a luminescent 2D material and color centers in diamond}

\author{Richard Nelz}
\author{Mariusz Radtke}
\author{Abdallah Slablab}
\affiliation{Universit\"at des Saarlandes, Fakult\"at NT, Physik, 66123 Saarbr\"ucken, Germany}%
\author{Zai-Quan Xu}
\author{Mehran Kianinia}
\author{Chi Li}
\author{Carlo Bradac}
\author{Igor Aharonovich}
\email{Igor.Aharonovich@uts.edu.au}
\affiliation{School of Mathematical and Physical Sciences, Faculty of Science, University of Technology Sydney, Ultimo, NSW, 2007, Australia}
\author{Elke Neu}
\email{elkeneu@physik.uni-saarland.de}
\affiliation{Universit\"at des Saarlandes, Fakult\"at NT, Physik, 66123 Saarbr\"ucken, Germany}%

\date{\today}


\begin{abstract}
Energy transfer between fluorescent probes lies at the heart of many applications ranging from bio-sensing and -imaging to enhanced photo-detection and light harvesting. In this work, we study F\"orster resonance energy transfer (FRET) between shallow defects in diamond---nitrogen-vacancy (NV) centers---and atomically-thin, two-dimensional materials---tungsten diselenide (WSe$_2$). By means of fluorescence lifetime imaging, we demonstrate the occurrence of FRET in the WSe$_2$/NV system. Further, we show that in the coupled system, NV centers provide an additional excitation pathway for WSe$_2$ photoluminescence. Our results constitute the first step towards the realization of hybrid quantum systems involving single-crystal diamond and two-dimensional materials that may lead to new strategies for studying and controlling spin transfer phenomena and spin valley physics.
\end{abstract}

\maketitle
\section{Introduction \label{sec:intro}}
F\"orster resonance energy transfer (FRET) is the near-field transfer of energy due to the dipole-dipole interaction in a donor/acceptor pair system. FRET transfers energy from an excited donor to an acceptor which is initially in the ground state. The transfer does not involve the exchange of photons and occurs at a rate determined by the overlap integral, i.e.\ the frequency-dependent oscillator strengths of the quantum spectroscopic (fluorescence/absorption) transition dipoles as well as their distance.\cite{Clegg2006} For the donor, FRET provides a new decay channel and consequently reduces its excited state lifetime. In contrast, for the acceptor FRET constitutes a non-radiative excitation pathway, thus not inducing lifetime changes but potentially enhancing acceptor luminescence.  
Realized for a variety of heterogeneous donor-acceptor systems---quantum dots, molecules, chromophores, etc.\cite{kim_energy_2016}---FRET constitutes the basis of a wide range of both fundamental effects and practical applications including single-molecule (bio)sensing and (bio)imaging, super-resolution fluorescence microscopy, as well as FRET-enhanced photodetection and light harvesting.

Evidently, the characteristics of the donor/acceptor systems are key to the practical realization of any FRET-based application. As fluorescent dye molecules traditionally used in FRET may bleach, recent work has focused on exploring resonant energy transfer processes in luminescent solid-state systems. These show superior stability and much broader versatility for many of the proposed sensing, imaging and optoelectronic applications. Luminescent point defects in diamond,\cite{Aharonovich2014a} including the well-studied nitrogen vacancy (NV) center,\cite{doherty_nitrogen-vacancy_2013} are long-term stable, fluorescent probes with established nanoscale sensing capabilities for magnetic\cite{Rondin2014} and electric fields,\cite{Dolde2014} as well as temperature.\cite{Neumann2013} Furthermore, FRET processes involving NV centers in nanodiamonds have already been demonstrated with organic molecules \cite{Mohan2010a,Tisler2011} and graphene.\cite{Tisler2013a,brenneis2015ultrafast} 
Simultaneously, there is an entire family of fluorescent probes in atomically-thin two-dimensional (2D) transition metal dichalcogenide (TMDs) materials which exhibit ultra-bright luminescence. 2D-TMDs are leading candidates for emerging applications in optoelectronics, photodetection and valleytronics \cite{schaibley2016valleytronics,mak2016photonics,peng2015two} rendering them significant for quantum technologies.\cite{toth2019single} So far, FRET involving luminescent 2D-TMDs has been demonstrated with organic dye molecules \cite{Zhou2019} and colloidal quantum dots.\cite{Zang2016, Prins2014} Notably, electrical gating of 2D-TMDs enables control over the efficiency of the energy transfer process.\cite{Prasai2015} It is noteworthy that FRET has been straightforwardly established between systems of different dimensionality (e.g.\ 1D: molecules, quantum dots, color centers; 2D:TMDs) and involving dipole transitions of fundamentally different nature (atom-like transitions in color centers and molecules, excitons in quantum dots and TMDs) \cite{Zhou2019,Zang2016, Prins2014, Tisler2013a, Prasai2015}. While the FRET mechanism itself is quite universal, the distance dependence of the process will reveal the nature of the participating dipoles. Conventionally, the rate of energy transfer between a pair of point dipoles such as molecules and chromophores scales with the distance $z$ between the dipoles like $\frac{1}{z^6}$. In contrast, FRET between an exciton delocalized in 2D, as observed in graphene or in our case in WSe$_2$, and a quantum dot or an atomic-scale quantum system like a color center scales with $\frac{1}{z^4}$.\cite{Goodfellow2014, Tisler2013a} 
\begin{figure}[h!]
	\centering
	\includegraphics[width=0.5\linewidth]{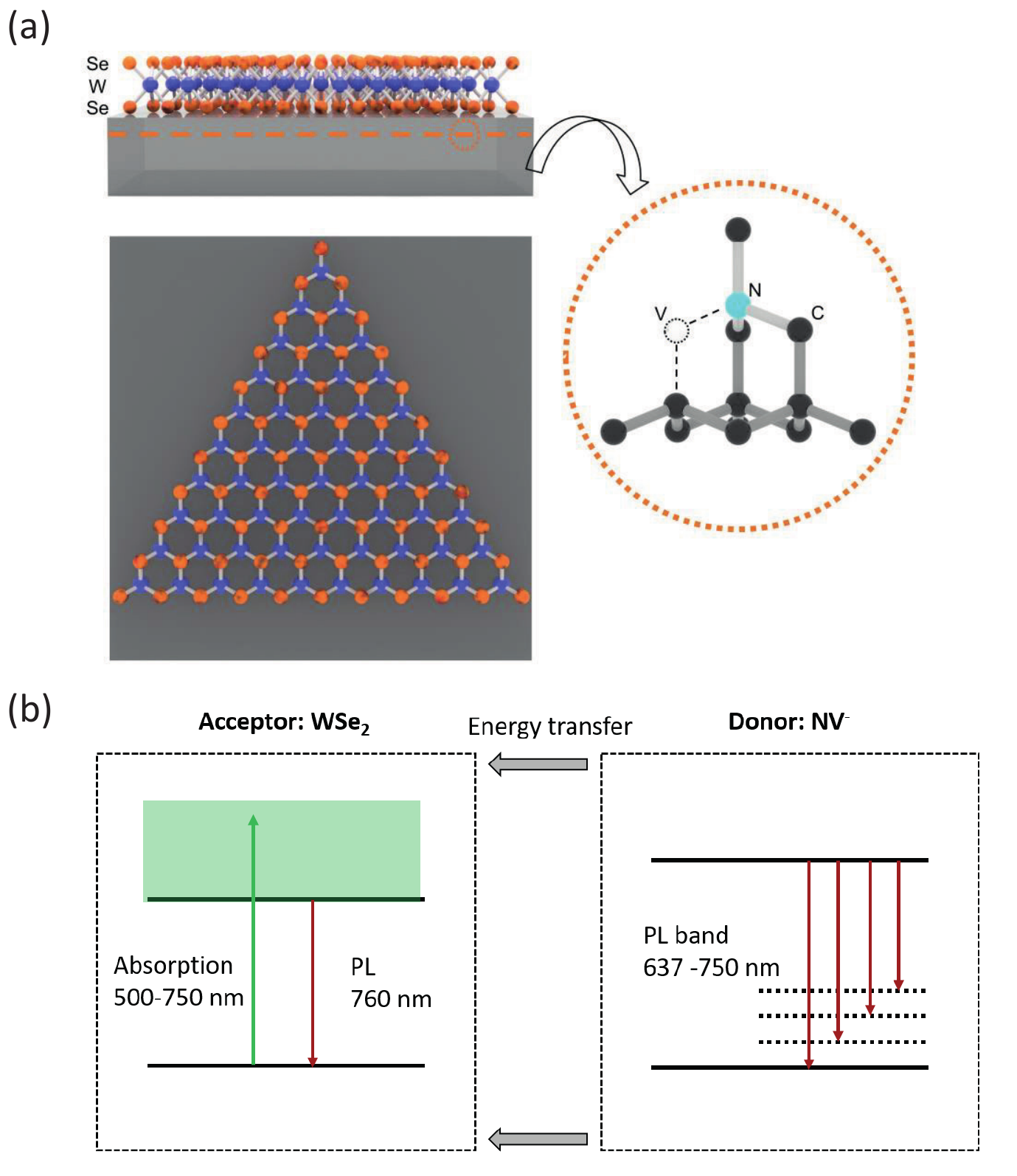}
	\caption{(a) Schematics of the NV/WSe$_2$ hybrid system illustrating the crystal structure of WSe$_2$ as well as the structure of the NV color center. The transfer of WSe$_2$ flakes to the SCD surface is described in Sec.\ \ref{sec:sampleprep}. (b) Schematics of the FRET process investigated here: negative NV centers serve as donors, monolayer WSe$_2$ flakes as the acceptor in the FRET process. The strong overlap of the absorption band of the WSe$_2$ flakes with the emission band of the NV centers establishes the main pre-requisite for FRET.}
	\label{fig:FRETscheme}
\end{figure}
In this work, we report FRET between two solid-state, stable, luminescent quantum systems---an ensemble of shallowly implanted NV centers in single-crystal diamond (SCD) and a 2D WSe$_2$ monolayer. The investigated hybrid system is depicted in Fig.\ \ref{fig:FRETscheme}(a). Although the observation of FRET between NV centers and molecular donors or 2D materials like graphene has already been demonstrated for nanodiamonds \cite{Tisler2011,Tisler2013a}, this is the first demonstration of FRET for shallowly implanted NV centers in SCD. In the process, NV centers act as donor dipoles which non-radiatively transfer their excitation energy to excitons in WSe$_2$. WSe$_2$ is an optimal FRET partner for NV centers, as its broad absorption band (500--750 nm) \cite{niu2018thickness,huang2013large} largely overlaps with the NV photoluminescence (PL) band between 640 and 750 nm. Figure \ref{fig:FRETscheme}(b) established the schematics of the donor/acceptor pair NV/WSe$_2$.  We employ fluorescence lifetime imaging to measure lifetime changes due to FRET between WSe$_2$ flakes and NV centers. We also observe enhanced excitation of WSe$_2$ via FRET processes. We estimate the FRET radius for the NV-WSe$_2$ pair to be 13 nm, and we show that the spin-based magnetic sensing capabilities of NV centers are conserved when FRET takes place. 

\section{Sample Preparation and experimental setup \label{sec:sampleprep}}


We use high-purity, (100)-oriented, synthetic, single-crystal diamond (SCD) from Element Six (electronic grade quality, [N]$^s$ $<$ 5 ppb, B $<$ 1 ppb). The SCD sample (size $2\times4$ mm$^2$) is polished to a roughness of R$_a < 3$ nm by Delaware Diamond Knives. We first employ reactive ion etching\cite{Challier2018} to remove the top 15 $\mu$m of the SCD. We thus avoid creating NV centers in the diamond top layers which are potentially damaged as a result of the mechanical polishing. We then form a homogeneous layer of NV centers by shallowly implanting nitrogen ions with an implantation density of $4\times10^{11}$ (nitrogen ions)/cm$^2$ and an energy of 4 keV. During the implantation, the sample is tilted by 7$^{\circ}$ with respect to the ion beam to avoid ion channeling. The SCD sample is then annealed in vacuum at 800 $^\circ$C and cleaned in boiling acids (1:1:1 mixture of sulfuric acid, perchloric acid and nitric acid). Using Monte Carlo simulations,\cite{SRIM} we estimate a resulting depth of the NV centers of (6.5 $\pm$ 2.7) nm below the SCD surface. As the geometry of our implantation process avoids ion channeling, SRIM simulations are a valid approximation for the NV depth distribution in our case. Using PL measurements, we show the creation of a spatially homogeneous ensemble of NV centers. Assuming typical creation yield in the order of 1\%, the above given implantation density leads to a sparse NV ensemble with $<10$ NV centers in the focus of the confocal microscope. We also employ reactive ion etching onto a target area of the SCD to remove selectively all NV centers (area size $\approx$ $1\times0.4$ mm$^2$). Using this area, we characterize the properties of our WSe$_2$ flakes on SCD independently of interactions with NV centers. 

\begin{figure}[h!]
	\centering
	\includegraphics[width=1\linewidth]{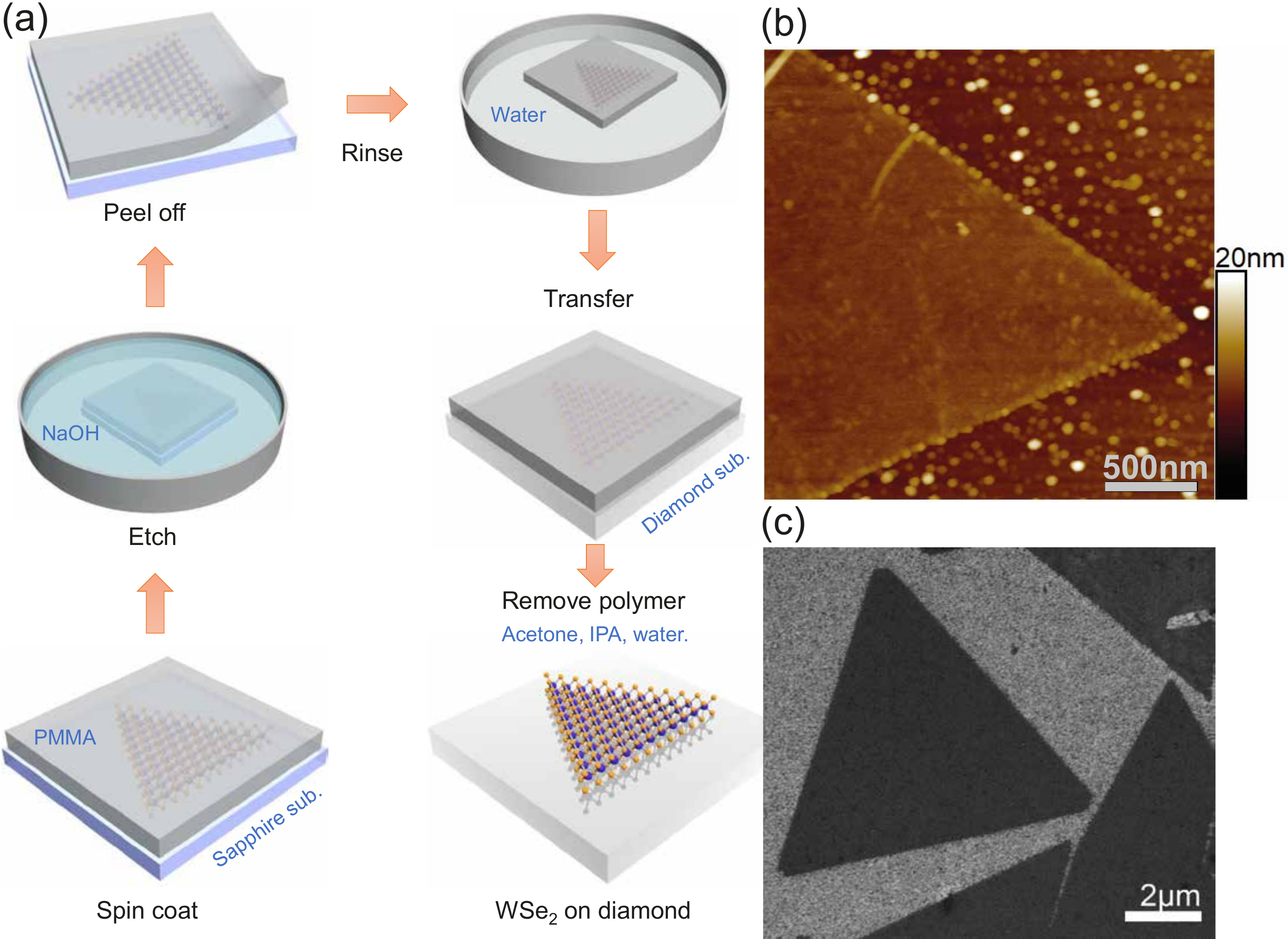}
	\caption{(a) Transfer process of the WSe$_2$ flakes onto the SCD surface as described in the main text. (b) Atomic force microscope image of a WSe$_2$ flake on top of a SiO$_2$ substrate. (c) Scanning electron microscope image of WSe$_2$ flakes on a SCD surface.}
	\label{fig:WSe2}
\end{figure}
We synthesize WSe$_2$ monolayer flakes on a sapphire (0001) substrate via chemical vapor deposition following Ref.\ \cite{huang2013large}. We transfer the flakes  onto our SCD sample using the method described below and depicted in Fig.\  \ref{fig:WSe2}(a). We first spin coat poly(methyl methacrylate) (PMMA; Microchem, A4, MW 495K, solvent: anisole) at 3,000 rpm for 1 minute onto the sapphire substrate with the flakes. We subsequently cure the sample by heating it up on a hotplate at 180 $^{\circ}$C for 1 minute. We then immerse the PMMA-coated substrate in 2M NaOH solution at 80 $^{\circ}$C for 1 hour to reduce the adhesion between the PMMA and the sapphire substrate. Subsequently, we peel off the PMMA layer together with the WSe$_2$ flakes exploiting the surface tension between them. The PMMA/WSe$_2$ composite layer is rinsed three times in deionized (DI) water and placed onto the clean SCD surface (or onto a SiO$_2$ substrate for further characterization, see below). Heating on a hot plate for 20 minutes at 150 $^{\circ}$C enhances the adhesion of the WSe$_2$ flakes to the new substrate. Finally, we immerse the SCD sample in hot acetone for 10 minutes and proceed by flush washing it in isopropanol (IPA) and DI water to remove the PMMA layer. The method has a high transfer yield over a mm-sized area, as confirmed by the density of WSe$_2$ flakes on the SCD being similar to that on the sapphire before transfer. The WSe$_2$ flakes mostly retain their triangular shape after transfer onto different substrates [see Fig.\ \ref{fig:WSe2}(b) and (c)]. We use atomic force microscopy to measure the thickness of our WSe$_2$ flakes. For a reference WSe$_2$ flake transferred onto a SiO$_2$ substrate, we find a thickness of 1.3 nm and a root mean square (rms) roughness of 0.4 nm. We note that this transferred flake is 0.6 nm thicker than our as grown-flakes. We attribute this observation to intercalated water in-between the flake and SCD as well as potential residuals from the transfer on top of the WSe$_2$ flake. We emphasize that the strong PL of the WSe$_2$ flakes at 760 nm  presented in Sec.\ \ref{sec:PLandlifetime} together with the AFM-based thickness measurements clearly identify the flakes as monolayers. 

A custom-built confocal scanning microscope (numerical aperture 0.8, pinhole size 50 $\mu$m) was used to investigate the interaction between the NV centers in the SCD and the WSe$_2$ monolayer. We use a tuneable (450--850 nm), pulsed laser (NKT EXW-12, pulse length $\approx$ 50 ps) equipped with a filter system (NKT SuperK Varia) as excitation source enabling pulsed PL and lifetime measurements. To acquire standard confocal microscopy PL maps, we excite the sample continuous diode-pumped solid-state laser with a wavelength of 532 nm and collect the detected PL signal through a 650 nm longpass filter. The collected PL signal was either sent to highly-efficient photon counters (Excelitas SPCM-AQRH-14, quantum efficiency $\approx$ 70 \%) or to a grating spectrometer (Acton Spectra Pro 2500, Pixis 256OE CCD). We employ a single-photon counting time correlator (PicoQuant, PicoHarp 300) to perform time-resolved PL analysis. The measured instrument response function (IRF) of our setup is modelled by a Gaussian function with a full width at half maximum (FWHM) of 326 ps. When fitting PL decay curves, we use exponential decays convoluted with our IRF and consequently correct the obtained PL lifetimes for the IRF. When performing PL lifetime imaging of NV centers, we filter the PL in a spectral window of 680-720 nm. We note that consequently, our NV lifetime measurements do not reflect contributions from neutral NV centers but only from negative NV centers. The setup is equipped with a microwave source (Stanford Research Systems, SG384) and an amplifier (Mini Circuits, ZHL-42W+) which allow for the delivery of microwaves through a 20 $\mu$m thick copper wire to enable electronic spin manipulation for NV centers. Table \ref{tab:expparameters} summarizes all important experimental parameters.
\begin{table}[h!]
    \centering
\begin{tabular}{|l|l|l|}
\hline
	& NV$^-$ & WSe$_2$ \\ 
	\hline 
	Absorption (nm)     & 460-630          & 500-750\cite{huang2013large}\\ 
	Emitted PL (nm)     & 637-750          & 760$\pm$20        \\
	depth  (nm)& 6.5$\pm$ 2.7     &            \\
	thickness (nm)&      & 1.3(4)            \\
	Detection (nm)      & 680-720          & $>$650            \\
	Excitation (nm)     & cw: 532          & cw: 532           \\
	                    & pulsed: 460-600  & pulsed: 460-600    \\
	\hline
\end{tabular}
    \caption{Experimental parameters at a glance.}
    \label{tab:expparameters}
\end{table}

\section{Photoluminescence and lifetime measurements \label{sec:PLandlifetime}}
\begin{figure}[h!]
	\centering
	\includegraphics[width=1\linewidth]{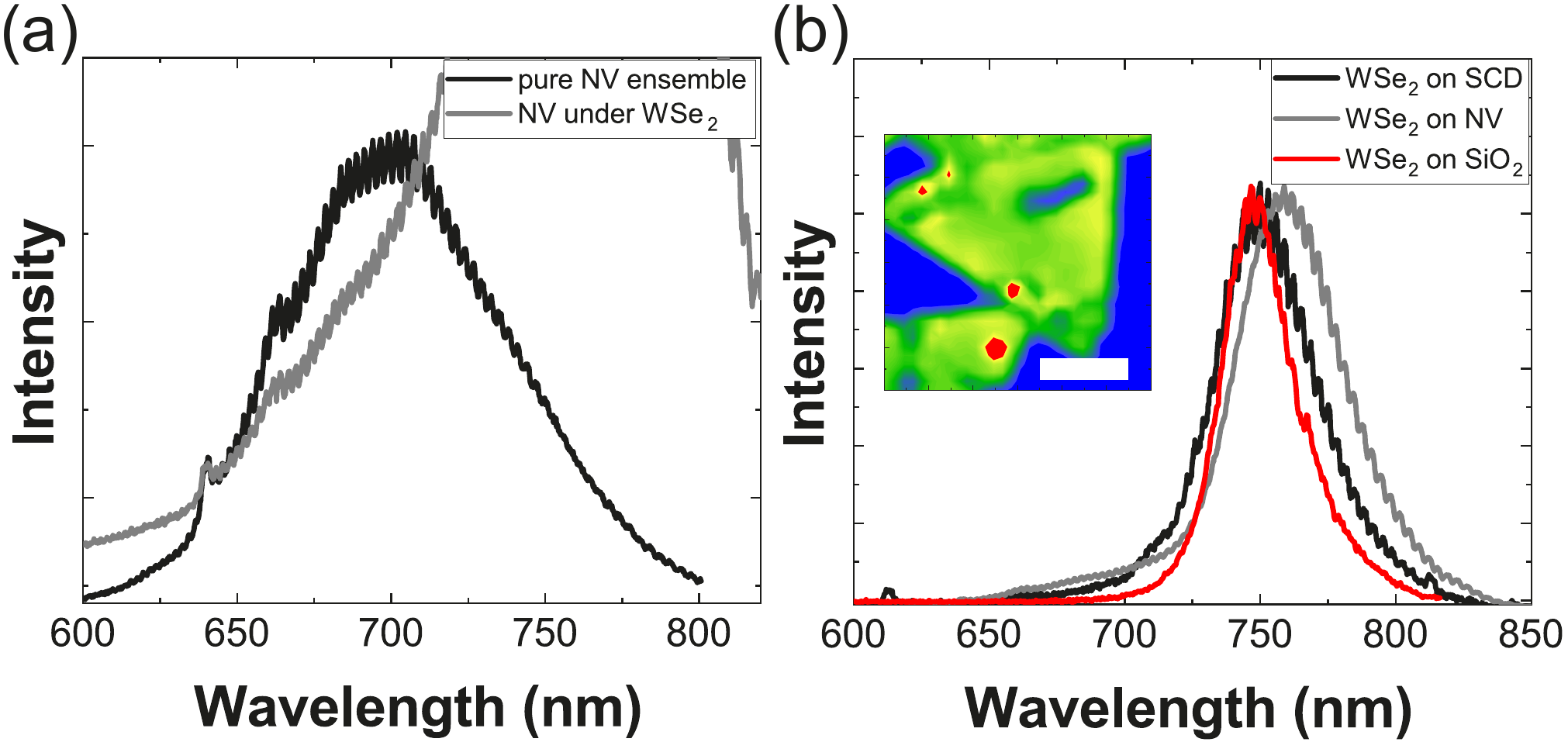}
	\caption{PL spectrum of the NV ensemble measured between WSe$_2$ flakes [black curve in (a)] showing the NV zero-phonon-line (at 637 nm) and phonon sideband. Gray curves in (a) and (b) show the PL spectrum of a WSe$_2$ flake with underlying NV ensemble. In (a) we normalize both spectra to the NV ZPL to facilitate comparison. In (b), we compare WSe$_2$ PL spectra on bare SCD (black) and SiO$2$ (red) substrates as well as on the NV center ensemble in SCD (gray). All spectra have been normalized to their peak value. All spectra clearly reveal the characteristic PL of WSe$_2$ monolayers centered around $\sim$ 760 nm with maximum peak shifts of $\sim$ 10 nm.  In contrast, the PL of a bilayer would occur centered around $\approx$ 815 nm.\cite{huang2013large} The inset shows a PL map (scale bar 2 $\mu$m) of a transferred WSe$_2$ flake (bright area) on top of the NV center ensemble (blue) recorded detecting wavelengths above 650 nm using continuous laser excitation at 532 nm (exc. power 700 $\mu$W).}
	\label{fig:ScanSpec}
\end{figure}

First, we characterize the hybrid NV/WSe$_2$ system depicted in Fig.\ \ref{fig:FRETscheme}(a) under continuous excitation. The inset in Fig.\ \ref{fig:ScanSpec}(b) shows a typical PL map of a triangular WSe$_2$ flake which we localize due to its strong excitonic PL at 760 nm [see Fig.\ \ref{fig:ScanSpec}(b), gray curve]. We find very similar PL spectra for our WSe$_2$ flakes transferred onto SiO$_2$ [see Fig.\ \ref{fig:ScanSpec}(b), red curve] and on the bare SCD [see Fig.\ \ref{fig:ScanSpec}(b), red curve]. The observed PL is characteristic for WSe$_2$ monolayers as previously reported.\cite{huang2013large} Despite the dominant WSe$_2$ PL signal, we identify a zero-phonon line at 637 nm originating from NV centers located underneath the flakes [Fig.\ \ref{fig:ScanSpec}(a), gray curve]. As expected, in-between individual WSe$_2$ flakes, we only observe PL due to implanted NV centers. We note that for both cases, NV ensemble below WSe$_2$ flakes as well as bare NV ensemble, we do not observe any PL due to neutral NV centers. We consequently infer that the presence of WSe$_2$ does not significantly alter the NV charge state.

\begin{figure}[h!]
	\centering
		\includegraphics[width=0.7\linewidth]{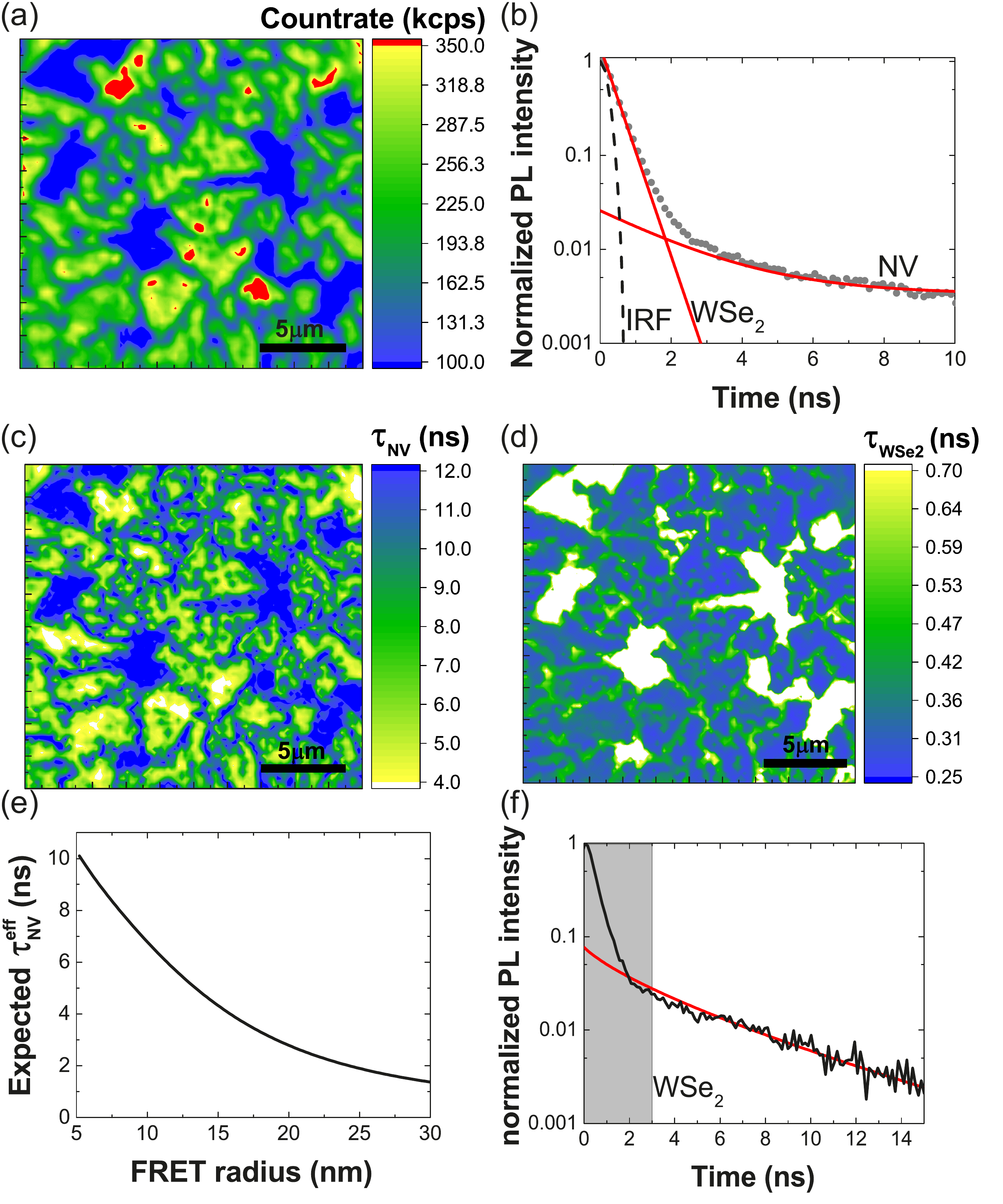}
	\caption{(a) PL map of WSe$_2$ flakes on NV ensemble. Here we use pulsed excitation ($\sim$ 9 nJ, 8 MHz repetition rate, 2 s integration time per pixel, excitation bandwidth 530--534 nm). (b) Exemplary lifetime measurement of the flakes and the underlying NV centers. The contributions of the instrument response (IRF), the WSe$_2$ PL and the NV center PL are drawn separately. (c), (d) Lifetime maps using $\tau_{NV}$ and $\tau_{WSe_2}$ corresponding to the PL map in (a).
	We consistently observe quenching of NV centers underneath the WSe$_2$ flakes with $\tau_{NV}<$ 6 ns. In contrast, between the flakes, we consistently find $\tau_{NV}^{bulk}$ and a lower overall PL level [blue areas in (a) and (c)]. As we only detect NV center PL in these areas, we fit the corresponding data with a mono-exponential decay. Consequently, in these areas, no values for $\tau_{WSe_2}$ are obtained which we represent in (d) in white color. (e) Observable NV ensemble lifetimes $\tau_{NV}^{eff}$ extracted from the simulation as a function of the FRET radius $R$. From our measurements, we extract a FRET radius of $R = $ 13 nm. (f) Simulated PL decay (red line) for the NV center ensemble using a F\"orster radius $R = $ 13 nm in comparison to the observed PL decay (black line). Model and observed data agree very well.}
	\label{fig:FRETRadius}
\end{figure}

To further investigate the hybrid NV/WSe$_2$ system, we perform PL lifetime measurements. In the area in which all NV centers have been removed, we find a lifetime of the WSe$_2$ exciton recombination PL of $\tau_{WSe_2}=0.41(5)$ ns. This is consistent with $\tau_{WSe_2}$ for pristine WSe$_2$ flakes before transfer. For our NV centers as donors, FRET will provide a new, additional decay channel. Consequently, their excited state lifetime ($\tau_{NV}$) will reduce. In contrast, for the acceptor, FRET establishes a (non-radiative) excitation pathway. Thus FRET is not expected to change $\tau_{WSe_2}$ which is in accordance with our observations for pristine flakes, flakes on SCD not coupled to NV centers and flakes coupled to NV centers [Fig.\ \ref{fig:FRETRadius}(d) and discussion below].

Figure \ref{fig:FRETRadius}(b) shows the PL decay recorded on a WSe$_2$ flake placed on the SCD surface with the shallowly-implanted ensemble of NV centers. We fit two main components: a fast decay with a time constant of 0.42(3) ns and a slower decay with a time constant of 5.1(3) ns. While the fast decay clearly arises due to the WSe$_2$ PL ($\tau_{WSe_2}$), we attribute the slower decay to the NV ensemble interacting with the WSe$_2$ flake ($\tau_{NV}$). In contrast, in areas not covered by WSe$_2$ flakes we consistently measure a much longer NV lifetime with an average value of $\tau_{NV}^{bulk}\sim$ 12(1) ns [see Fig.\ \ref{fig:FRETRadius}(c)], typical for NV centers in bulk diamond.\cite{Manson2006} For shallowly implanted NVs, slightly longer lifetimes of $\sim$ 16--17 ns have been previously reported.\cite{Nelz2019} We point out that, in our experiment, residuals from the transfer process might slightly reduce the NV lifetime in-between the flakes. Excitation with short ($<$ ns) laser pulses potentially reveals internal spin dynamics of NV centers including spin dependent lifetimes .\cite{Robledo2011a} The here observed value of $\tau_{NV}^{bulk}\sim$ 12(1) ns corresponds to the lifetime expected for the m$_s$=0 state.\cite{Robledo2011a} Indeed the high repetition rate (8 MHz) pulsed excitation that we use here will polarize the NV centers into m$_s$=0 as the time interval between subsequent pulses is several orders of magnitude shorter than typical NV T$_1$ times.     
For NV centers coupled to WSe$_2$ flakes, we find that $\tau_{NV}$ is halved compared to the non-coupled NVs in our sample. This finding proves the occurrence of non-radiative energy transfer---FRET---between NV centers in SCD and excitons in the WSe$_2$ flake. The WSe$_2$ flakes provides the NV centers with a non-radiative decay channel---mediated by dipole-dipole interaction---which reduces $\tau_{NV}$.\\

To further investigate FRET between WSe$_2$ and NV centers, we perform PL lifetime imaging of different areas of the SCD sample [see Fig.\ \ref{fig:FRETRadius}(a), (c) and (d)]. We fit a double exponential decay to the measured data and extract $\tau_{NV}$ and $\tau_{WSe_2}$. We consistently observe $\tau_{NV}<6$ ns in areas where the SCD surface is covered by a WSe$_2$ flake, as discernible from comparing the PL map in Fig.\ \ref{fig:FRETRadius}(a) and the lifetime map in Fig.\ \ref{fig:FRETRadius}(c). In contrast, we find $\tau_{NV}^{bulk}$ on all other positions. The pattern of the WSe$_2$ flakes is furthermore confirmed when plotting $\tau_{WSe_2}$ [see Fig.\ \ref{fig:FRETRadius}(d)].

Below, we interpret these results in detail and highlight the peculiarities of the FRET process between an ensemble of NV centers and a WSe$_2$ flake. In our case, the FRET process is non-trivial as FRET strongly depends on the distance between the donor and the acceptor and our NV centers show a spread of depths $z$ in the SCD. Assuming the flake is in direct contact with the SCD surface, each specific NV center of the ensemble lies at a different distance $z$ from the WSe$_2$. We thus would expect a complicated multi-exponential PL decay corresponding to a spread of $\tau_{NV}$. Experimentally, however, we find that the NV PL decay is very well described by a single exponential decay with time constant $\tau_{NV}$. To simulate the PL decay expected from the NV ensemble coupled to the WSe$_2$ flake, we first calculate $\tau_{NV}$ as a function of $z$. To this end, we need to determine the non-radiative decay rate $\gamma_{non-rad}$ due to the FRET process. We assume that the exciton is fully delocalized in the WSe$_2$ flake and the flakes have infinite size compared to the atomic-sized NV centers; we find \cite{Tisler2013a}

\begin{equation}\label{equ:GammaNR}
	\gamma_{non-rad}(z) = \gamma_{rad} \frac{R^4}{z^4}
\end{equation}
where $R$ is the F\"orster radius, i.e. the distance at which the efficiency of the FRET mechanism is 50 \%. The quantity $R$ depends on the quantum efficiency of the donor and on the spectral overlap between the donor's emission and the acceptor's absorption, as well as their dipole moments. We furthermore assume that the radiative decay rate $\gamma_{rad}$ is constant for all NV centers and equal to the reference bulk value $(\tau_{NV}^{bulk})^{-1}$.
For each NV center at a specific depth $z$, we should find a mono-exponential decay with $\tau_{NV}(z)$. Equation \ref{equ:GammaNR} shows that FRET is more efficient if $z$ is smaller; consequently NV centers very close to the surface will be strongly quenched and emit less photons. We calculate the PL intensity $I(z)$ of an NV center at a depth $z$ as
\begin{equation}\label{equ:Intensity}
	I(z) = I_0 \frac{\gamma_{rad}}{\gamma_{rad}+\gamma_{non-rad}(z)}
\end{equation}
where $I_0$ is the non-quenched PL intensity, and $\gamma_{rad}$ and $\gamma_{non-rad}$ are the radiative and non-radiative decay rates, respectively.

To obtain the resulting PL decay of the ensemble, we weight each of the mono-exponential decay curves for each depth $z$ with the intensity $I(z)$ (Eq.\ \ref{equ:Intensity}) and with the depth distribution $D(z)$ of NV centers resulting from the implantation process. We extract the depth distribution $D(z)$ using Monte Carlo Simulations (SRIM), which produce a depth profile of (6.5 $\pm$ 2.7) nm. As we observe all NV centers in the ensemble simultaneously, we integrate over the whole implantation profile to retrieve the observed PL decay.

\begin{figure}[h!]
	\centering
	\includegraphics[width=1\linewidth]{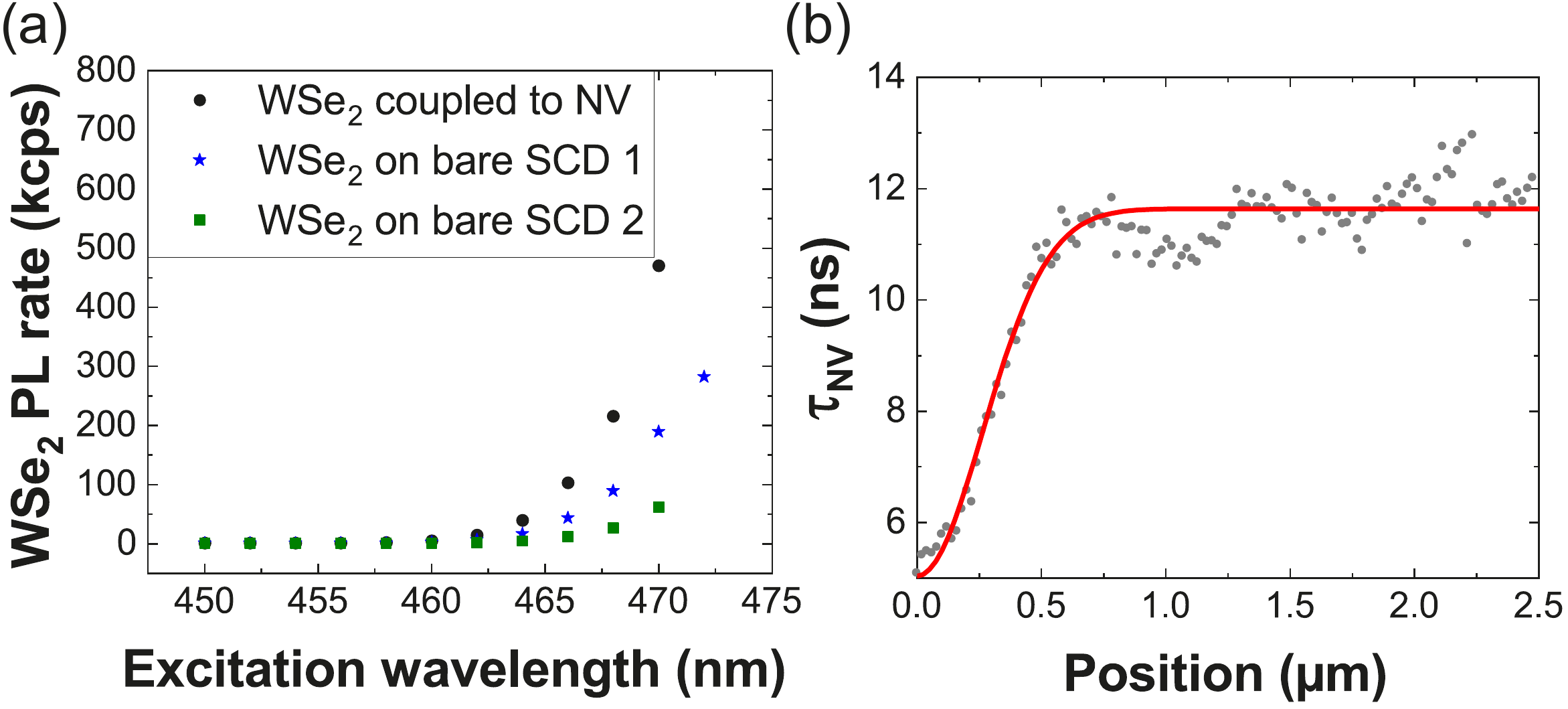}
	\caption{(a) Median PL rate of the WSe$_2$ flakes for different excitation wavelengths $\lambda_{exc}$. Black dots: PL of WSe$_2$ coupled to NV centers. WSe$_2$ PL rates are corrected for background as well as for the PL of the underlying NV ensemble. blue stars and green squares: PL of WSe$_2$ on bare SCD as recorded in two areas (1,2) separated by more than 0.5 mm.  These measurements demonstrate that FRET from NV centers constitutes an additional excitation pathway for  WSe$_2$ PL (details see text). (b) NV lifetime recorded along a line perpendicularly crossing the edge of a WSe$_2$ flake. We fit a Gaussian to the transition from bulk lifetime to the quenched lifetime (red curve). We obtain a FWHM of 620(30) nm which is comparable to the point spread function measured for our confocal setup.}
	\label{fig:FRETProof}
\end{figure}

Notably, the calculated PL decay of the NV ensemble indicates a decay which is well-represented with a mono-exponential function with an effective $\tau_{NV}^{eff}$ (assuming $R >$ 5 nm) in full agreement with our experimental observations. Figure \ref{fig:FRETRadius}(e) shows the expected value of $\tau_{NV}^{eff}$ for 5 nm $< R <$ 30 nm---which we now use to estimate $R$ for the NV/WSe$_2$ pair and find $R_{NV/WSe_2}=$ 13 nm. To further confirm the agreement between the measured data and our model, we reduce the data to the time range in which the NV's mono-exponential decay is dominating. We do not include data from the first 3 ns---as this signal mainly represent PL from WSe$_2$---and we do not consider the long time tail---i.e. all the data points with count rates below 1 \% of the NV centers' peak value. This data treatment leads to the black line in Fig.\ \ref{fig:FRETRadius}(f) which agrees very well with the calculated PL decay (red line), assuming $R_{NV/WSe_2}=$ 13 nm, and resulting in $\tau_{NV}^{eff} =$ 5.2 ns. We note that $\tau_{NV}=$ 5.2 ns in this situation is expected for NV centers in a depth of $\sim$ 12 nm. Our value for $R_{NV/WSe_2}$ is comparable to that reported for the NV-graphene system, $R_{NV/graphene}$,\cite{Tisler2013a} strongly supporting the hereby observation of FRET between NV centers and WSe$_2$.\\
To further investigate the energy transfer mechanism between NV centers and WSe$_2$ flakes, we study how the WSe$_2$ PL intensity depends on the excitation wavelength $\lambda_{exc}$. Negatively-charged NV centers act as donors for WSe$_2$. Consequently, FRET from excited NV centers constitutes an excitation path for WSe$_2$ PL which adds to laser excitation of WSe$_2$ PL. Especially when the direct laser excitation of the acceptor is inefficient, FRET enhances the PL of the acceptor (and simultaneously reduces the PL of the donor). Recording PL spectra of ensembles of NVs at positions not covered by WSe$_2$ flakes, we clearly observe PL from negatively-charged NV centers for $\lambda_{exc} > 465$ nm. We however note that we cannot infer a ratio of neutral and negative NV centers for excitation with blue laser light as the employed setup does not permit observing spectral features due to neutral NV centers (in contrast to the setup we used for $\lambda_{exc}$= 532 nm, see Fig.\ \ref{fig:ScanSpec}).  However, as we clearly observe negatively-charged NV centers, they will contribute via FRET to the excitation of WSe$_2$ PL for $\lambda_{exc} > 465$ nm. We thus investigate this excitation wavelength range in detail. We investigate the PL rate of the WSe$_2$ flakes as a function of $\lambda_{exc}$ [see Fig.\ \ref{fig:FRETProof}(a)] and aim to compare the situation where WSe$_2$ flakes are placed on the NV ensemble and the situation in which the WSe$_2$ flakes are placed on SCD regions where all NV centers have been removed via etching (see Section II). To account for variations in the PL intensity for different flakes, we compare the median count-rate of WSe$_2$ in areas of 20 by 20 $\mu$m$^2$. To check for consistency, we furthermore investigate two areas without NV centers separated by more than 0.5 mm. We also correct the WSe$_2$ PL intensity for NV center and background PL. As we compare the WSe$_2$ PL intensity with and without NV centers, changes in the excitation laser power when changing $\lambda_{exc}$ affect both measurements in the same way. Consequently, we did not correct for variations of the laser power as well as a potentially wavelength dependent transmission of our setup. We note that as consequence of this, our measurements do not necessarily reveal the wavelength dependent PL excitation probability of NV centers.  Figure \ref{fig:FRETProof}(a) displays a clear tendency of an enhanced excitation of WSe$_2$ PL for flakes coupled to the shallow NV ensemble. We note that the enhanced excitation cannot be due to the absorption of NV PL by the WSe$_2$ flakes for the PL is too weak to induce the observed enhancement. The observed excitation enhancement of WSe$_2$ in the presence of NV centers constitutes additional strong evidence for FRET between NV centers and WSe$_2$.
\\

\begin{figure}[h!]
	\centering
	\includegraphics[width=1\linewidth]{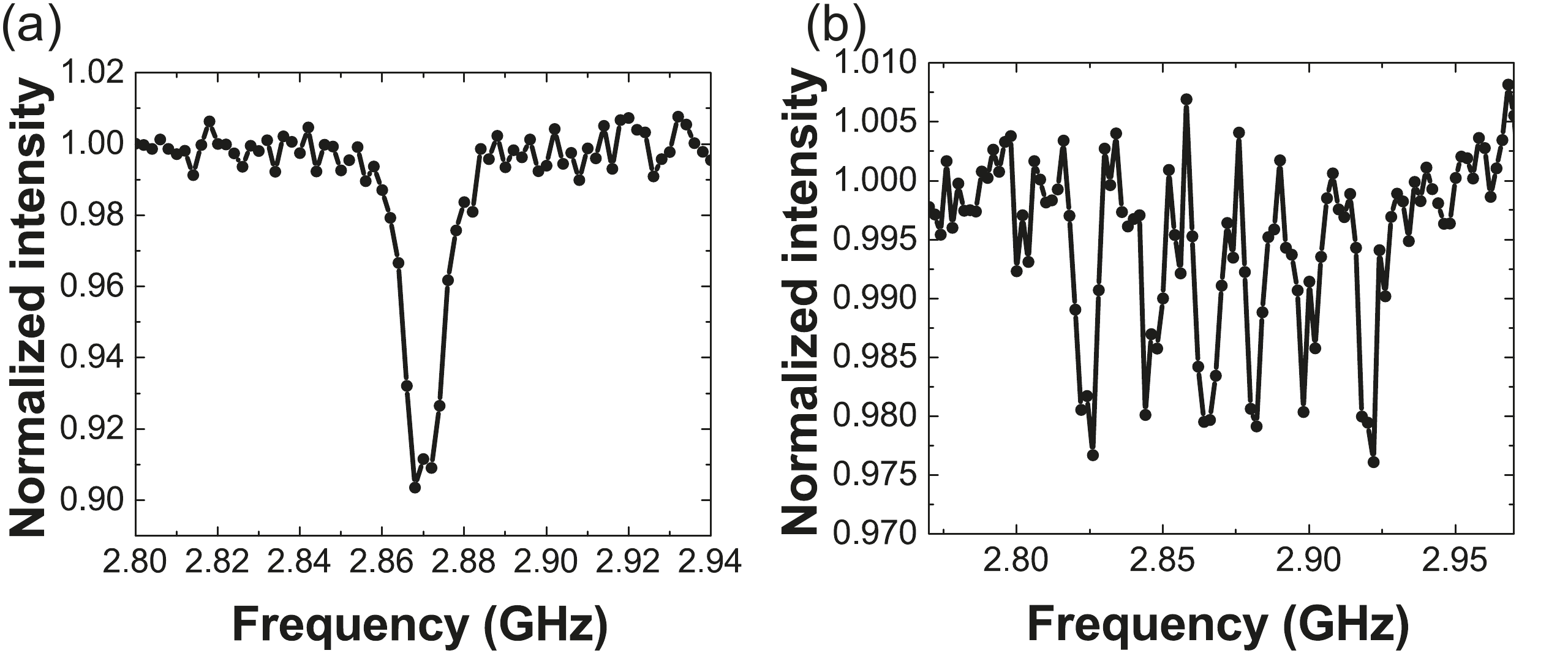}
	\caption{Optically detected magnetic resonance of the nitrogen vacancy centers ensemble without (a) and with an external magnetic field of 25 G applied (b) indicating the potential for NV centers as multi-functional sensors.}
	\label{fig:MultiSensor}
\end{figure}

We now investigate the electronic spin properties for NV centers under the WSe$_2$ flakes. We use lifetime-gating to separate NV PL and WSe$_2$ PL to enhance the measurement contrast when performing optically detected magnetic resonance (ODMR) measurements under pulsed laser excitation. Fig.\ \ref{fig:MultiSensor}(a) and (b) show the ODMR of the NV ensemble underneath the WSe$_2$ flake in the absence and presence of an external magnetic field, respectively. We observe an ODMR contrast of 10 \% without external field, which is typical for shallow NV centers, and which proves clear separation of NV and WSe$_2$ PL. Observing ODMR with a magnetic-field-dependent splitting for NV centers undergoing FRET indicates that they can serve as multi-functional sensors: while using FRET processes to monitor the presence of other dipoles, NV centers can simultaneously sense magnetic fields. This observation renders NV centers promising as multi-functional sensors in biological systems or for the investigation of novel materials where they can operate as nanoscale probes for nuclear magnetic resonance spectroscopy \cite{Lovchinsky2017} that simultaneously couple to excitons via FRET.\\
We finally investigate how precisely we can localize the edge of a WSe$_2$ flake using the spatial variation of $\tau_{NV}$. We measure $\tau_{NV}$ along a line perpendicularly crossing the edge of a WSe$_2$ flake [see Fig.\ \ref{fig:FRETProof}(b)]. We fit $\tau_{NV}$ using a Gaussian function approximating the point spread function of our setup. We find a FWHM of 620 nm which is closely matching the point-spread-function (PSF) of our setup which we estimated by imaging single color centers in nanodiamonds. Consequently, localizing the edge of a WSe$_2$ flake is limited by the setup's PSF only indicating the possibility for high resolution imaging using the here investigated FRET process. We now address the imaging speed of the PL lifetime imaging. Typical detected PL rates of our NV ensemble amount to 70 kcps for a pulse energy of $\sim$ 10 nJ ($\lambda_{exc}=$ 530-534 nm, Rep.\ Rate 8 MHz) Following Ref.\ \cite{Gratton2003}, we estimate the minimum number of photons needed to reliably determine $\tau_{NV}$ to be 1000. This leads to a minimum integration time per pixel of $<$ 15 ms.\\

\section{Summary and Outlook}
In conclusion, we have demonstrated FRET between shallow NV centers in SCD and WSe$_2$ flakes with an estimated F\"orster radius of 13 nm. The FRET process strongly reduces $\tau_{NV}$ to around 6 ns, whereas the coupling to the NV centers enhances the excitation of WSe$_2$ for $\lambda_{exc}$ below 500 nm. We show that NVs undergoing FRET retain their ODMR and are applicable as multi-functional sensors.

In the future, we will investigate the transfer of WSe$_2$ flakes onto SCD photonic structures e.g.\ nanopillars with single NV centers. The first tests conducted during this work prove that typical SCD nanopillars are robust in the applied transfer process. Using such photonic structures will enhance the PL rates from single NV centers and will also allow for the modification of the excitonic properties of 2D materials via inducing local strain.\cite{palacios2017large}
While traditionally FRET pairs are formed by attaching FRET partners to larger molecules or nanoparticles \cite{Loura2011,Mohan2010a} or directly within a biological specimen \cite{Loura2011}, the extension to stable solid-state systems could enable the realization of scanning devices where FRET is established between a single quantum probe scanning the system under investigation. Consequently, the distance between sample and probe can be varied continuously, which allows for in-depth characterization of the FRET process and imaging the sample on the nanoscale. Such techniques termed FRET-Scanning near field optical microscopy (FRET-SNOM) \cite{Sekatskii2008} will highly-profit from stable probes like NV centers in SCD. The first demonstration of FRET-SNOM enabled nanoscale imaging of graphene flakes using a scanning NV center in a nanodiamond.\cite{Tisler2013a} Moreover, hybrid systems involving NV centers and 2D materials are potential candidates for spin transfer and spin valley physics. The latter has triggered intense research in TMDs and has potential for quantum information and sensing applications.\cite{Lu2019} The here investigated FRET process can potentially aid to determine the depth of shallow color centers in SCD given the FRET radius of the WSe$_2$/NV pair is precisely known. One advantage of this approach would be its applicability to color centers that do not posses an electronic spin. \\

This research has been funded via the NanoMatFutur program of the German Ministry of Education and Research (BMBF) under grant number FKZ13N13547. The Australian research council (via DP180100077, DP190101058 and DE180100810), the Asian Office of Aerospace Research and Development Grant FA2386-17-1-4064, and the Office of Naval Research Global under Grant N62909-18-1-2025 are gratefully acknowledged. We acknowledge funding via a PostDoc fellowship of the Daimler and Benz foundation. We thank J\"org Schmauch for recording SEM images as well as Dr.\,Rene Hensel for use of the ICP-RIE and Michel Challier for help with the RIE etching of the sample.

\bibliographystyle{ChemEurJ}
\bibliography{Literaturverzeichnisaktuell_UTS}
\end{document}